\documentclass[12pt, a4paper]{article}
\usepackage[latin1]{inputenc}
\usepackage[dvips]{graphicx,color}
\usepackage{amsmath}
\usepackage{amsfonts}
\usepackage{amssymb}
\numberwithin{equation}{section}
\begin{document}
\def\pplogo{\vbox{\kern-\headheight\kern -29pt
\halign{##&##\hfil\cr&{\ppnumber}\cr\rule{0pt}{2.5ex}&\ppdate\cr}}}
\makeatletter
\def\ps@firstpage{\ps@empty \def\@oddhead{\hss\pplogo}%
  \let\@evenhead\@oddhead 
}
\def\maketitle{\par
 \begingroup
 \def\thefootnote{\fnsymbol{footnote}}
 \def\@makefnmark{\hbox{$^{\@thefnmark}$\hss}}
 \if@twocolumn
 \twocolumn[\@maketitle]
 \else \newpage
 \global\@topnum\z@ \@maketitle \fi\thispagestyle{firstpage}\@thanks
 \endgroup
 \setcounter{footnote}{0}
 \let\maketitle\relax
 \let\@maketitle\relax
 \gdef\@thanks{}\gdef\@author{}\gdef\@title{}\let\thanks\relax}
\makeatother
\def\eq#1{Eq. (\ref{eq:#1})}
\newcommand{\nc}{\newcommand}
\def\theequation{\thesection.\arabic{equation}}
\nc{\beq}{\begin{equation}}
\nc{\eeq}{\end{equation}}
\nc{\barray}{\begin{eqnarray}}
\nc{\earray}{\end{eqnarray}}
\nc{\barrayn}{\begin{eqnarray*}}
\nc{\earrayn}{\end{eqnarray*}}
\nc{\bcenter}{\begin{center}}
\nc{\ecenter}{\end{center}}
\nc{\ket}[1]{| #1 \rangle}
\nc{\bra}[1]{\langle #1 |}
\nc{\0}{\ket{0}}
\nc{\mc}{\mathcal}
\nc{\etal}{{\em et al}}
\nc{\GeV}{\mbox{GeV}}
\nc{\er}[1]{(\ref{eq:#1})}
\nc{\onehalf}{\frac{1}{2}}
\nc{\partialbar}{\bar{\partial}}
\nc{\psit}{\widetilde{\psi}}
\nc{\Tr}{\mbox{Tr}}
\nc{\tc}{\tilde c}
\nc{\tk}{\tilde K}
\nc{\tv}{\tilde V}
\nc{\CN}{{\mathcal N}}
%

\setcounter{page}0
\def\ppnumber{\vbox{\baselineskip14pt
}}
\def\ppdate{RUNHETC-07-05} \date{}

\author{Michael R. Douglas$^{1,\&}$, Jessie Shelton$^1$
 and Gonzalo Torroba$^1$\\
[7mm]
{\normalsize $^1$NHETC and Department of Physics and Astronomy}\\
{\normalsize Rutgers University}\\
{\normalsize Piscataway, NJ 08855--0849, USA}\\
{\normalsize $^\&$I.H.E.S., Le Bois-Marie, Bures-sur-Yvette, 91440 France}\\
{\normalsize {\tt mrd, jshelton, torrobag@physics.rutgers.edu}}}

\title{\bf \LARGE Warping and Supersymmetry Breaking}
\maketitle \vskip 1cm

\begin{abstract} \normalsize
\noindent We analyze supersymmetry breaking by anti-self-dual
flux in the deformed conifold.  This theory has been argued to
be a dual realization of susy breaking by antibranes.  As such,
one might expect it to lead to a hierarchically small breaking
scale, but only if the warp factor is taken into account.
We verify this by explicitly computing the warp-modified moduli
space metric.  This leads to a new term,
with a power-like divergence at the conifold point, which
lowers the breaking scale.  We finally point out various puzzles
regarding the gauge theory interpretation of these results.

\end{abstract}

\bigskip
\newpage

\tableofcontents

\vskip 1cm
\section{Introduction}\label{sec:intro}

Over the last few years there has been much progress in
understanding flux compactifications of string theory
\cite{kdreview}.  Supersymmetric configurations have been well
studied~\cite{gkp,2b, 2a}, and as a general rule one can understand
their physics using either supergravity, or a dual gauge theory
picture in which fluxes are replaced by branes.  As a basic
example, consider pure super Yang-Mills theory.  Its vacuum
structure can be usefully described by the Veneziano-Yankielowicz
effective superpotential, which leads to an exponentially small
gaugino condensate.  The same physics can be described by a dual
theory with fluxes near a conifold singularity, and an analysis
based on the flux superpotential \cite{gkp}.

The situation for supersymmetry breaking configurations is much less
clear.  While there are many known examples of gauge theories in which
the dynamical scale sets the scale of supersymmetry breaking, these
always seem to involve special matter content or tuning in the
superpotential, and until fairly recently this evidence seemed to
suggest that in a ``generic'' set of theories (such as would come
out of an ensemble of string vacua), very few would break supersymmetry.

More recently, this belief has been significantly revised in light
of new work incorporating ingredients such as metastability,
``retrofitting,'' obstructions in the dual geometry, and many
others \cite{iss,dine,franco, ag,Giveon:2007fk, program},
suggesting that a far broader and more generic class of gauge
theories will break supersymmetry.

It should be said however that, so far, the constructions which
produce low scales still require either tuning of parameters or
discrete symmetries, so the question of what distribution of
supersymmetry breaking scales we expect to come out of string theory
remains open.  To illustrate this, consider the problem of finding
string theory realizations of the ``retrofitting'' construction of
\cite{dine}, in which a small parameter in the superpotential (for
example the small quark mass required in \cite{iss}) is obtained as
the dynamical scale of a second gauge group.  While one can easily
find superpotential couplings such as $(\Tr W_\alpha^2) (\tilde Q)^2$
which do this, the harder problem is to find a mechanism which
suppresses an order one bare quark mass term $(\tilde Q)^2$.  The only
obvious candidate is a discrete R symmetry \cite{Dine:2006xt},
which appears to be even
less natural in flux vacua than the tuning we are hoping to explain
\cite{Dine:2005gz}.

One might imagine that a truly generic construction would work both at
weak coupling (gauge theory) and strong coupling (the flux dual).
On the flux side,
nonsupersymmetric flux compactifications also exist and appear to be
generic \cite{Saltman:2004sn,2b}.  However, both statistical
analyses \cite{douglasns} and more intuitive arguments
\cite{Dine:2005yq,Giudice:2006sn}
suggest that these favor a high scale of supersymmetry breaking.
Although flux superpotentials contain the ingredients needed to obtain
low scales, in the work so far these do
not lead to a low supersymmetry breaking scale
without tuning.

In exploring these issues, we tried to find the simplest possible
flux compactifications with susy breaking.  One simple and popular
idea is to break supersymmetry by combining branes and antibranes,
which preserve incompatible $\mathcal N=1$ subalgebras of an
underlying $\mathcal N=2$ supersymmetry algebra of type II
Calabi-Yau compactification. This includes the anti D3-brane in
a conifold throat worked out in \cite{Kachru:2002gs},
a controlled construction using wrapped
D5-branes to get D-term breaking
in \cite{Diaconescu:2006nk}, a suspended brane
construction in \cite{Giveon:2007fk}, and many others.

In \cite{ag}, a system of D5 and anti-D5 branes wrapped on a pair
of resolved conifolds was studied, and argued to have a simple
flux dual involving imaginary anti-self-dual flux.  On the other
hand, as we will review, an $\mathcal N=1$ effective field theory
analysis of this theory along the lines of \cite{ag} leads to high
scale breaking.  While this may at first appear paradoxical,
actually it is to be expected, as the scale of susy breaking in
this model is the anti-D-brane tension, of order the string scale.

Of course, as has been much discussed in the string
compactification literature, starting with \cite{Kachru:2003aw},
supersymmetry breaking by antibranes can
lead to low scale breaking, but only if the antibranes are localized
in a region of large warp factor.  Now in the limit considered by
\cite{ag}, in which $\alpha'\rightarrow 0$ before considering other
effects, the warp factor is not present, since it is sourced by fluxes
which are quantized in units of $\alpha'$.  Thus their results appear
self-consistent, but this suggests that we need to incorporate the
warp factor to see low scale susy breaking.

In this work, we will do just this.  By a detailed analysis of warping
effects on the metric on complex structure moduli space, we show the
existence of a term with a power-like divergence close to the conifold
point.  This contribution will be the one responsible for lowering the
scale of the breaking.

Having explained our basic results, the order of the discussion will
be reversed, for clarity.  We begin in section \ref{sec:susybreak}
by reviewing the anti-D5 brane theory and its flux dual.  In section
\ref{sec:warpcomp} we
develop the general theory of warped Calabi-Yau compactification,
along lines initiated by Giddings and collaborators \cite{gid1,gid2}.
In particular, we analyze general properties of the warped
metric $G_{\alpha \bar \beta}$ for complex deformations. We show
that throats contribute new zero modes to the warp factor, that
are responsible for enhancing $G_{\alpha \bar \beta}$ near
conifold points in moduli space. Since the general analysis leaves
numerical factors undetermined, in section \ref{sec:explconif} we
compute explicitly $G_{\alpha \bar \beta}$ for the deformed
conifold. This analysis reveals a new $|S|^{-4/3}$ contribution.

In subsection ~\ref{subsec:closer} we use this to understand the
supergravity behavior near the tip of the warped deformed conifold. We
also show how a negative $F_3$ flux through the 3-cycle of the
conifold, and a far away $O7$ plane preserving a misaligned
supersymmetry, can give parametrically small supersymmetry breaking.
This still requires potentially non-generic ingredients, depending on
choices made in the bulk, so the question of whether this mechanism
generically leads to low scale breaking remains open.

Finally, section \ref{sec:gauge} comments on the dual gauge
theory description, in which anti $D5$s wrap the resolved cycle.
While the new term $|S|^{-4/3}$ actually matches with old expectations for the
K\"ahler potential for the gaugino condensate,
we are left with numerous unanswered questions here.

\section{Supersymmetry breaking without warping} \label{sec:susybreak}

We begin by reviewing the suggestion of \cite{ag} that the dual
of an anti-D5-brane wrapped on the resolved conifold is an
anti-self-dual flux in the deformed conifold geometry, leading to
supersymmetry breaking.  However, using the unwarped moduli
space metric, we find that this is high scale breaking.

\subsection{Flux compactification of IIb string}

We start with $\mathcal N=2$ compactification of type IIb string
theory, in which the complex structure moduli live in vector
multiplets.  We choose  a symplectic basis $(A_\alpha, B^\alpha)$
of $H_3(X,\mathbb Z)$, and take as coordinates on moduli space the
A cycle periods $\{(S^\alpha)\}$. The B cycle periods can then be
integrated to define a prepotential $\mathcal F$, so that
\begin{equation} \label{eq:complexdef}
S^\alpha=\int_{A_\alpha} \Omega\;,\;\,\frac{\partial \mathcal
F}{\partial S^\alpha}=\int_{B^\alpha} \Omega\,.
\end{equation}
The metric on moduli space can then be determined either from
Calabi-Yau geometry, or from the prepotential, as
\begin{equation} \label{eq:Gnowarp}
G_{\alpha \bar \beta}= {\rm Im}\,\partial_\alpha\partial_{\bar \beta} {\mathcal F}
= -\frac{\int \chi_\alpha \wedge
\chi_{\bar \beta}}{\int \Omega \wedge \overline \Omega}
\end{equation}
where  $\chi_\alpha$, $\alpha=1,\ldots, h^{2,1}$ are a basis of
$H^{2,1}(X, \mathbb C)$.

We next consider a compactification with
three-form fluxes $G_3:=F_3-\tau H_3$. We define the
flux parameters
$(N_R^\alpha, N_{NS}^\alpha, \beta_\alpha^R, \beta_\alpha^{NS})$
by
\begin{equation} \label{eq:generalfluxes}
4 \pi^2 \alpha'
\int_{A_\alpha} G_3=N_R ^{\alpha}- \tau
N^{\alpha}_{NS}\;,\;\,
-4 \pi^2 \alpha'
\int_{B^\alpha} G_3=\beta^R _{\alpha}- \tau
\beta_{\alpha}^{NS}\; .
\end{equation}
These are quantized in units of $4 \pi^2 \alpha'$,
which we generally set to one in the following.
We follow the notation in~\cite{gt}.

As discussed in \cite{gkp}, the vacuum energy of the fluxes
depends on the complex structure moduli and dilaton, leading
to moduli stabilization.  This can be analyzed by minimizing a
scalar potential derived from the Gukov-Taylor-Vafa-Witten
superpotential
\begin{equation} \label{eq:GVW}
W=\int G_3 \wedge \Omega\, ,
\end{equation}
using the standard $\mathcal N=1$ supergravity formalism.

We should say from the start that this $\mc N=1$ effective supergravity
description in terms of the Calabi-Yau moduli fields is in general
only a very partial description of the physics.  For example, there
might be other light modes, from KK modes in string compactification,
other degrees of freedom in the gauge theory dual, and the like.  This will
be even more true once we take warping into account, as discussed in
\cite{burges}.  We do not intend to study this question in detail here, but
rather will limit ourselves to finding an effective potential which properly
describes the vacuum energy, moduli stabilization and supersymmetry breaking.
At least on the strongly coupled (supergravity) side, we believe our arguments
(based on explicit $d=10$ solutions) suffice to demonstrate this.

\subsection{Geometric properties}\label{subsec:geometry}

For completeness we include a quick review of the basic
properties of the conifold.
The algebraic variety describing the deformed conifold is
\begin{equation} \label{eq:defconif}
u^2+v^2+y^2-x^2+S=0\,.
\end{equation}
Its holomorphic properties are very simple. There are only two
nontrivial 3-cycles, $(A, B)$, $A \cap B =1$; for $S \to 0$, $A
\to 0$ and $B$ is noncompact. The $A$ cycle is an $S^2$ fibration
($u,\, v\; \in \mathbb R$) over the cut $x \in (-\sqrt S,\,+\sqrt
S)$ of the hyperelliptic curve
\begin{equation} \label{eq:ellipt}
F(x,y)=y^2-x^2+S=0\,.
\end{equation}
The noncompact $B$-cycle extends between $y=\pm \infty$ and runs
through the previous cut. We introduce a geometrical cutoff
$\Lambda_0$ such that the points at infinity become $\pm
\Lambda_0$. From the usual monodromy arguments, the periods are
\begin{equation} \label{eq:conifperiod}
\int_A \Omega=S\;,\;\,\int_B \Omega=\frac{\partial \mathcal
F}{\partial S}=\Pi_0 + \frac{1}{2\pi i}\,S\,{\rm
log}\,\frac{\Lambda_0^3}{S}+\ldots
\end{equation}
where $\ldots$ are analytic subleading contributions and $\mathcal F$ is the
prepotential of the geometry.

\subsection{SUGRA on the warped deformed conifold} \label{subsec:sugradef}

A simple embedding of the conifold in a compact
Calabi-Yau orientifold was constructed in~\cite{gkp}.
However the details
of the embedding do not matter for our discussion, so
we consider the following simplified model. We `zoom in' to
a local neighborhood of a compact CY X, containing the deformed
conifold (\ref{eq:defconif}).
The only information that we keep
from the rest of $X$ is that there is an orientifold projection,
breaking $\mathcal N=2 \to \mathcal N=1$, and preserving the
chiral supermultiplet with scalar component $S$.
While a complete
discussion requires solving the D3 tadpole condition, which
usually requires adding wandering $D3$ branes, we keep these
away from the conifold, so that they don't enter these results.

In the presence of the 4-form $C_4$, compactifying on the conifold
contributes an $\mathcal N=2$ 4d vector multiplet $\mathcal A=(S,
\psi, \lambda, A_\mu)$, where $S$ is the complex modulus of the
conifold
$$
\frac{\partial g_{\bar i \bar j}}{\partial S}=-\frac{1}{\parallel
\Omega \parallel^2}\,S(x)\,\bar \Omega_{\bar i}^{\phantom{1}kl}
\chi_{S\,kl\bar j}\,.
$$
and $A_\mu$ is the $U(1)$ gauge field from $C_4=A_\mu(x) dx^\mu
\wedge \chi_S$. $\psi$ and $\lambda$ are the fermion
superpartners.

The orientifold action is~\cite{louis}
$$
\mathcal O = (-)^{F_L}\Omega_p \sigma^*\;,\;\sigma^*
\Omega=-\Omega\,;
$$
$\Omega_p$ is the worldsheet parity, $F_L$ is the left moving 4d
fermion number and $\sigma^*$ is the holomorphic involution
(acting on forms). This will produce $O3/O7$ planes.
Orientifolding splits the $\mathcal N=2$ vector
multiplet into an $\mathcal N=1$ chiral multiplet $(S, \psi)$ and
a vector multiplet $(\lambda, A_\mu)$. Since we want to keep $S$
as a low energy 4d field, we take the action of the involution to
be $\sigma^* \chi_S=-\chi_S$. In this way the vector multiplet is
projected out and we are left with only $(S, \psi)$.

As the next step we turn on the following quantized fluxes:
\begin{equation} \label{eq:fluxconif}
\int_A F_3=N\;,\;-\int_B F_3=\beta^R\;,\;-\int_B H_3=\beta^{NS}\,.
\end{equation}
Because of the monodromy $B \to B+n A$, $\beta^R$ is defined ${\rm
mod}\, N$, and will play the role of a discrete $\theta$-angle. In an $\mathcal N=2$
formalism, fluxes may be seen as FI terms for the auxiliary components of the
superfield $\mathcal A$. Based on this identification, it
was noted in~\cite{ag} that for $N >0$ (we always take
$\beta^{NS}>0$), the supersymmetry variations are
$$
\delta_{\epsilon} \psi=0\;,\,\;\delta_{\epsilon} \lambda = i \epsilon \,\frac{1}{{\rm Im}\,
\partial^2_S \mathcal F}\,\big(\frac{i}{g_s} \beta^{NS}+N \overline{\partial^2_S \mathcal F} \big)\,.
$$
Therefore positive flux respects the same supersymmetry as the
orientifold; we still have an $\mathcal N=1$ theory because
$\lambda$ is projected out from the spectrum, so $\delta \lambda
\neq 0$ is not seen.

Using (\ref{eq:conifperiod}) and (\ref{eq:fluxconif}), the GVW
superpotential (\ref{eq:GVW}) for the conifold reads
\begin{equation} \label{eq:superpconif}
W=\frac{N}{2\pi i}S\big({\rm log}\frac{\Lambda_0^3}{S}+1
\big)-\big(\beta^R-\frac{i}{g_s}\beta^{NS} \big)S\,.
\end{equation}
For $N >0$, $\beta^{NS}>0$, solving $\partial_S W=0$,
\begin{equation} \label{eq:Ssusy}
S=e^{-2\pi i \beta^R/N}\,e^{-2\pi \beta^{NS}/g_sN}\,\Lambda_0^3\,.
\end{equation}
There are $N$ degenerate vacua coming from $\beta^R=0, \ldots,
N-1$; to simplify our results, we will in general set $\beta^R=0$
and remember this degeneracy. The supergravity background with
fluxes corresponds to the warped deformed conifold, appropriately
glued into the compact CY, as discussed in section
\ref{subsec:gluing}.

This type of flux dual was used in the discussion of supersymmetry breaking
by anti-D3 branes in \cite{Kachru:2002gs}.

\subsection{Supersymmetry breaking}\label{subsec:susydef}

We now consider the effect of misaligning the supersymmetry
preserved by the $O7$ and the one preserved in the conifold, by
turning on negative flux $N<0$.

In the case $N<0$, Aganagic \etal\ %
\cite{ag} showed that $\delta \lambda=0$ but
\begin{equation}\label{eq:dpsi}
\delta_{\epsilon} \psi = i \epsilon \,\frac{1}{{\rm Im}\,
\partial^2_S \mathcal F}\,\big(\frac{i}{g_s} \beta^{NS}
 +N \partial^2_S \mathcal F \big)\,.
\end{equation}
and hence this flux configuration breaks $\mathcal N=1 \to
\mathcal N=0$ spontaneously.

For $N <0$, $\beta^{NS}>0$, (\ref{eq:Ssusy}) would give a result
$S \gg \Lambda_0^3\,$! It was argued in~\cite{ag} that the
physical vacuum is instead the minimum of the  scalar
potential
$$
V = e^K \left( G^{i\bar i} \partial_i W {\bar\partial}_{\bar i} W^* \right) ,
$$
obtained from \eq{superpconif} and \eq{Gnowarp}.  This is located at
\begin{equation} \label{eq:Snsusy}
S_{N<0}=e^{-2\pi i \beta^R/N}\,e^{-2\pi
|\beta^{NS}/g_sN|}\,\Lambda_0^3\,.
\end{equation}
On this vacuum,
\begin{equation} \label{eq:Fterm}
\partial_S W_{N<0}=2i \frac{\beta^{NS}}{g_s} \neq 0
\end{equation}
and hence supersymmetry is broken by an explicit non-zero
F-term $\partial_S W$.

\subsection{Scale of supersymmetry breaking}

In principle there are various ways one could define this scale,
and in \cite{ag} definitions involving the mass splittings among
supermultiplets were studied.

However, in a spontaneously broken $\mathcal N=1$ supergravity
theory, the standard definition is the norm of the $F$ terms, or
equivalently the scale determined by the $F$ term contribution to
the scalar potential
$$
M_{susy}^4 = V = e^K \left( G^{i\bar i} D_i W D_{\bar i} W^* \right).
$$
In a realistic compactification with near-zero
cosmological constant, this scale
will also determine the gravitino mass, as
$m_{3/2}=M_{susy}^2/\sqrt{3}M_{Planck}$.  How exactly it enters into
observable susy breaking depends on the mediation mechanism, but
very generally one expects soft terms of order $M_{susy}^2/M_{Planck}$
from gravitational couplings and gravitino loop effects.
Thus one generally requires
$M_{susy} < 10^{11} \GeV$ (the intermediate scale) for a model which
naturally solves the hierarchy problem, and this is the operational
definition of a low scale of susy breaking.

Besides the superpotential \eq{GVW}
and the K\"ahler potential $K$,
one also needs the dependence on the K\"ahler moduli of $X$ to get
the full scalar potential.
Before taking into account stringy and quantum corrections, this leads to
``no scale'' structure.  This suffices to define $M_{susy}$ as above,
and we will discuss the nonperturbative effects later.

The factor $e^K$ also includes normalization factors determined by
doing the dimensional reduction, leading to~\cite{gid1}
\begin{equation} \label{eq:V2}
V=\frac{1}{2\kappa_4^2}\,\frac{1}{V_W {\rm Im}\,\tau ({\rm
Im}\,\rho)^3}\frac{1}{\parallel \Omega \parallel^2 V_W}\,G^{S \bar
S} |\partial_S W|^2\,.
\end{equation}

Using \eq{conifperiod} and \eq{Gnowarp}, we find
$$
G_{S\bar S} \sim c \log \frac{\Lambda_0^3}{|S|} .
$$
and thus
\begin{equation} \label{eq:V3a}
V =\frac{1}{2\kappa_{10}^2}\,\frac{g_s}{({\rm
Im}\,\rho)^3}\,\Big[c\, {\log}\,\frac{\Lambda_0^3}{|S|}\Big]^{-1}
 \big|\frac{N}{2\pi i}\,
 {\log}\,\frac{\Lambda_0^3}{S}+i \frac{\beta^{NS}}{g_S}\big|^2\;
\sim N\,\frac{\beta^{NS}}{g_s} .
\end{equation}

Since we are working in conventions in which $\alpha'$ is order
one, the upshot is that $\mathcal N=1$ supersymmetry is broken at
a high scale. This can be confirmed by a $d=10$ computation of the
mixing between the gravitino and the goldstino, here the fermionic
component of $S$.  The essential content of this computation is
already present in the supersymmetry variation \eq{dpsi}.

Since the energy \eq{V3a} is the expected tension of $N$ anti
D5-branes, in retrospect this result should not be very surprising.
However it raises the question of whether and how it would be
changed by including the warp factor.

\section{Warped Compactifications} \label{sec:warpcomp}

We start by reviewing the basic features of warped
compactifications, and then we describe the warp effects on the
geometry of the complex moduli space. We will concentrate on the
complex moduli stabilization.

\subsection{Warping and fluxes}\label{subsec:warpnflux}

We mainly follow DeWolfe and Giddings
~\cite{gid1}. Starting from an underlying CY $X$
with metric $g_{mn}(y)$, turning on fluxes produces a warped
metric
\begin{equation} \label{eq:metric1}
ds^2=e^{2A(y)} \eta_{\mu \nu} dx^\mu dx^\nu+e^{-2A(y)}g_{mn}(y)
dy^m dy^n\;
\end{equation}
$m,n=1, \ldots, 6$. To avoid confusion, we want to stress that we
are not using the usual GKP notation $\tilde g_{mn}$ for the CY
metric, because we want to avoid tildes appearing in all our
formulas. Throughout the work we will rise and lower indices only
with respect to $g_{mn}$, so that the dependence on the warp
factor is always explicit.

A complete and consistent dimensional reduction is
very involved, containing subtle issues from KK modes,
compensators and warping contributions~\cite{gid2}. However, the
results that we will discuss in this work refer to the vacuum
structure of the theory, arising purely from the effective
potential.  Fortunately, this is free of many of these issues.

In \cite{gid1}, a Kaluza-Klein reduction is done from $d=10$ IIb
supergravity to a $d=4$, $\mathcal N=1$ effective supergravity,
taking into account the warp factor.  To get a self-consistent
$\alpha'\rightarrow 0$ limit in which the warp factor remains, one
simultaneously increases the flux $G_3$, keeping the flux
parameters $(N_R^\alpha, N_{NS}^\alpha, \beta_\alpha^R,
\beta_\alpha^{NS})$ fixed.  The other $\alpha'$ and quantum
corrections in the full string theory are dropped.

The result is that, in the presence of warping, the
superpotential still takes the form \eq{GVW}, but
the metric on complex structure moduli space
is deformed to
\begin{equation} \label{eq:Gwarp1}
G_{\alpha \bar \beta}=-\frac{\int e^{-4A} \chi_\alpha \wedge
\chi_{\bar \beta}}{\int e^{-4A} \Omega \wedge \overline \Omega}\, ,
\end{equation}
where the warp factor $e^{-4A}$ is determined by a supergravity
equation of motion.  In a supersymmetric background, this
is~\cite{gkp}
\begin{equation} \label{eq:sugrawarp}
-\nabla^2(e^{-4A})=\pm \frac{G_{mnp}\,\bar G^{mnp}}{12\,{\rm
Im}\,\tau}\pm 2\kappa_{10}^2T_3 \rho_3^{loc}
\end{equation}
where
\begin{equation}
\nabla^2:=\frac{1}{\sqrt g}
\partial_m \sqrt g g^{mn} \partial_n\,.
\end{equation}
Here the plus (minus) sign corresponds to ISD (IASD) fluxes and D3
(anti D3) brane charge, with the consequence that the r.h.s. of
Eq. (\ref{eq:sugrawarp}) is always positive, as it should in order
to get a positive definite metric. These signs are easy to
understand: denoting, as in~\cite{gkp}
$$
\tilde F_5=(1+*)[d\alpha(y)\wedge dx^0\wedge dx^1\wedge dx^2\wedge
dx^3]\,;
$$
$\alpha$ will of course depend on the sign of its sources (
3-fluxes or 3-branes). But the BPS-like condition of GKP, extended
to IASD fluxes also, is $e^{4A}=\pm \alpha$, with the result that
the warp factor only depends on the absolute value of the sources.

In this limit, the scalar potential takes the
no-scale form, cancelling $|W|^2$ against
$|D_{\rho}W|^2$, where $\rho$ are the K\"ahler moduli.
The scalar potential for the complex moduli and dilaton
then takes the standard $\mathcal N=1$ form
\begin{equation} \label{eq:potential1}
V=\frac{G^{\alpha \bar \beta} D_\alpha W \overline{D_\beta
W}}{\int e^{-4A} \Omega \wedge \overline \Omega} \,.
\end{equation}

\subsection{`Warped' geometry of the moduli space}\label{subsec:warpedgeom}

Implicit in the previous results is the claim that
the warp-deformed moduli space
metric (\ref{eq:Gwarp1}) is K\"ahler.
In ~\cite{gid1} it was
suggested that this metric may be derived from the K\"ahler potential
\begin{equation} \label{eq:K}
K(S, \bar S)=-{\rm log} \big(-i \int e^{-4A} \Omega \wedge
\overline \Omega \big)\, ,
\end{equation}
in other words that
$\partial_\alpha \partial_{\bar \beta} K=G_{\alpha \bar \beta}$.

Seeing this requires properly treating the zero mode of the warp
factor, which will be important when
we match the noncompact conifold solution onto a compact bulk manifold.
We start from
\begin{equation} \label{eq:griffiths}
\frac{\partial \Omega}{\partial S^\alpha}=k_\alpha \Omega +
\chi_\alpha\;,\;\,k_\alpha=\frac{\int e^{-4A}
\partial_\alpha\Omega \wedge \overline \Omega}{\int e^{-4A} \Omega
\wedge \overline \Omega}\,.
\end{equation}
One can check that
this expression is valid with or without the warp factor $e^{-4A}$
as after integration it cancels between numerator and denominator.
Therefore,
\begin{equation} \label{eq:delK}
\partial_\alpha K=-k_\alpha-\frac{\int (\partial_\alpha e^{-4A})\, \Omega \wedge \overline \Omega}{\int e^{-4A}
\Omega \wedge \overline \Omega}\,.
\end{equation}
Note that the usual special geometry relation $\partial_\alpha
K=-k_\alpha$ is not yet apparent.
After some algebra,
\def\oob{\Omega\wedge{\bar\Omega}}
\begin{eqnarray}  \label{eq:relationGK}
\partial_\alpha \partial_{\bar \beta} K=G_{\alpha \bar
\beta}&
+\frac{\big(\int
\oob\,\partial_\alpha e^{-4A}\big)(\int
\oob\,\partial_{\bar \beta}e^{-4A}\big)}
{\left(\int  \oob\,e^{-4A(y)}\right)^2}
 \\
\nonumber
&-\frac{\int \oob\,\partial_\alpha\partial_{\bar \beta}e^{-4A}}
{\int \oob\,e^{-4A(y)}} .
\end{eqnarray}
The consistency of $\mathcal N=1$ supergravity requires that
$\partial_\alpha \partial_{\bar \beta} K=G_{\alpha \bar \beta}$,
in other words that the second
and third terms on the right hand side vanish.  The easiest way
for this to happen is if
\begin{equation} \label{eq:fixedzero}
0 = \int \oob\, \partial_\alpha e^{-4A} ,
\end{equation}
{\it i.e.} the zero mode $c$ of the warp factor, defined by writing
\begin{equation} \label{eq:cwarp}
e^{-4A(y)}=c+e^{-4A_0(y)} ,
\end{equation}
is independent of variations of the complex structure.
This is sensible because in a true compactification,
the equation of motion
\eq{sugrawarp} does not
determine the zero mode of $e^{-4A}$.  Rather, the zero mode
enters into the overall volume as a K\"ahler modulus (this is before
bringing in nonperturbative effects depending on K\"ahler moduli).

Another way to define the zero mode is to use the relation
\begin{equation} \label{eq:detgOmega}
\sqrt{\det g} =
\frac{1}{3!}(g_{i\bar i} dy^i d{\bar y}^{\bar i})^3
=\frac{\Omega \wedge \overline \Omega}{\| \Omega \|^2}
=\frac{\Omega \wedge \overline \Omega}{\frac{1}{3!}
  g^{i\bar i}  g^{j\bar j}  g^{k\bar k}
 \Omega_{ijk} \bar \Omega_{\bar i\bar j\bar k}} .
\end{equation}

One can rewrite the Ricci flatness condition on $g$ as the
condition that the denominator $\|\Omega\|^2$ in this expression
is constant on the Calabi-Yau, and thus \eq{fixedzero} is equivalent
to the condition that
$$
0 = \int \sqrt{\det g} \,\partial_\alpha e^{-4A} .
$$
Given an initial choice of the zero mode, this condition
determines a unique solution of \eq{sugrawarp}

Note that it will not in general be the case that
$\partial_\alpha e^{-4A} = 0$ pointwise, only under the integral.
Similarly, even though the unwarped volume $\int \sqrt{\det g}$
can be defined to be independent of complex structure, the
functional form of the volume element $\sqrt{\det g}$ can vary.
Thus the actual warped volume, defined as
$$
V_W = \int \sqrt{\det g} \, e^{-4A} ,
$$
could in general depend on the complex structure moduli.

To summarize: in the absence of fluxes, compactifying on $X$ gives
$\mathcal N=2$ supersymmetry and correspondingly the complex
moduli space $\mathcal M$ is a special Kahler manifold. When we
turn on fluxes, $X$ gets a warp contribution and we break
$\mathcal N=2 \to \mathcal N=1$. $\mathcal M$ is still a Kahler
manifold but no longer special. The prepotential $\mathcal F$ is
not enough to specify all the geometrical data of the conformal
Calabi-Yau, and the information on the fluxes comes in through the
factor $e^{-4A}$.

\subsection{Zero modes of the warp factor}\label{subsec:zeromodes}

Evidently the $e^{-4A}$ warp factor plays a crucial role.
While in ten-dimensional terms, it is determined by \eq{sugrawarp},
as yet there is no simple four-dimensional ansatz for this factor.
Thus we will shortly need to
look at the details of the ten-dimensional solution.
However there is a simple argument which suggests what we are looking
for.

As we discussed earlier, the general solution for the warp factor is
\begin{equation} \label{eq:cwarp2}
e^{-4A(y)}=c+e^{-4A_0(y)} ,
\end{equation}
where $c$ is a free parameter related to the total warped volume.
For example, in the large volume limit we can take $c \to \infty$,
which allows one to identify~\cite{gid2}
\begin{equation} \label{eq:cvol}
V_{CY} \sim c^{3/2\,.}
\end{equation}

In this limit, the warp factor becomes irrelevant.  On the other hand,
for any fixed $c$, the factor $e^{-4A_0(y)}$ might become so large
at some $y$ that it cannot be neglected.  In this sense, the neglect
of the warp factor at large volume is an order of limits problem.

The large variation of the warp factor is due to the source tems in
\eq{sugrawarp}.  As an example, around a D3-brane, which is a delta
function source $c_2 \delta^6(r)$ at $r=0$, one has $e^{-4A} \sim
c_2/r^4$.  A non-zero flux, of either sign, while not localized, has a
similar effect.

While the details depend on $d=10$ physics, the general behavior
is governed by a `localized' (rapidly decaying) but non-constant zero mode
of the source-free equation, in other words the Laplacian on $e^{-4A}$.
For simplicity, let us consider this in a single throat
located (in some local coordinate system) at $r=0$, embedded in a
compact Calabi-Yau. The discussion extends without changes to any
number of throats.
Then, in a conifold geometry, the Laplacian takes the form
\begin{equation}
\frac{1}{r^5}\frac{\partial}{\partial r}\big(r^5
\frac{\partial}{\partial r} G\,\big)+\frac{1}{r^2}\nabla_{\Psi}^2
G=0\,,
\end{equation}
where $r$ is the conical distance from the singularity $r=0$, and
$\Psi$ denotes the five angular variables. Hence the zero mode with
trivial angular dependence on the throat is
\begin{equation} \label{eq:zeroA}
G(r)=c+\frac{c_2}{r^4}\,.
\end{equation}
The general solution (\ref{eq:cwarp2}) becomes
\begin{equation} \label{eq:Gwarp}
e^{-4A(y)}=c+\frac{c_2}{r^4}+e^{-4A_0(y)}\,.
\end{equation}

The zero mode term $c_2/r^4$ will then dominate other effects in the
throat. Expanding the Kahler potential in the
large volume limit,
$$
K(S, \bar S)= -{\rm log}\Big(-i \int_{bulk}e^{-4A} \Omega \wedge
\overline \Omega\,-i \int_{conif}e^{-4A} \Omega \wedge \overline
\Omega \,\Big)
$$
\begin{equation} \label{eq:Kwarp0}
\approx K_0+i e^{K_0} \int_{conif} (c+\frac{c_2}{r^4})\,\Omega
\wedge \overline \Omega\,,
\end{equation}
with $e^{-K_0} :=-i \int_{bulk} c \,\Omega \wedge \overline
\Omega$. This is essentially the limit from local special geometry
to rigid special geometry, valid in the neighborhood of the
conifold point in moduli space~\cite{gt}.

This shows that there can be a
warp enhancement of the Kahler potential at the tip of the throat.
Moreover, the metric computed from here will be
\begin{equation} \label{eq:Gwarp0}
G_{S \bar S} \approx i\,e^{K_0} \,\int_{conif}
\big(c+\frac{c_2}{r^4} \big) \chi_S \wedge \chi_{\bar S}\,,
\end{equation}
which agrees with the general result (\ref{eq:Gwarp1}). We will see
below that, at least for the deformed conifold, the
$(2,1)$ forms $\chi_\alpha$ are also localized at $r=0$,
contributing to the enhancement effect.

The coefficient $c_2$ is fixed
by a standard Stokes theorem argument.
Given a region $R \subset M$ with boundary $\partial R$, we
have\footnote{Recall that, from Eq.(\ref{eq:sugrawarp}), a plus
(minus) sign must be used in the e.o.m. of the warp factor for ISD
(IASD) fluxes and positive (negative) D3 charge.}
$$
\int_{\partial R} \sqrt{g_5} \,\partial_{\hat n} e^{-4A} =\int_R
\sqrt{g_6}\, \nabla^2 e^{-4A} = \pm \int_R \sqrt{g_6}
 \,\frac{G_{mnp}\,\bar G^{mnp}}{12\,{\rm Im}\,\tau}\pm 2\kappa_{10}^2T_3
$$
so the normal derivative of $e^{-4A}$ integrated over the boundary is
proportional to the total source contained in the region.  At large
$r$, the boundary integral is dominated by the contribution of the
zero mode $c_2/r^4$.

Thus, D3 charge near the conifold, including that induced by
fluxes, leads to warping. Of course this charge will be
compensated by negative charge elsewhere, say from O3 planes.

\section{Explicit analysis of the deformed conifold}\label{sec:explconif}

Let us see how this shows up in an explicit treatment. Thus,
following Klebanov and Strassler~\cite{ks}
we discuss the CY metric corresponding to
(\ref{eq:defconif}). We warm up with the singular conifold, with
$S=0$. It has
\begin{equation} \label{eq:singularg}
ds_6^2=dr^2+r^2\,ds_{T^{1,1}}^2\;,\;\,ds_{T^{1,1}}^2=\frac{1}{9}(g^5)^2+\frac{1}{6}
\sum_{i=1}^4 (g^i)^2\,.
\end{equation}
The basis of one forms $g^i$ was introduced
in~\cite{Minasian:1999tt,ks}; they
arise from the angular variables of the base $S^2\times S^3$. In
terms of
\begin{equation} \label{eq:singularforms}
\omega_2:=\frac{1}{2}(g^1 \wedge g^2+g^3 \wedge
g^4)\;,\;\,\omega_3:=\frac{1}{2}g^5\wedge(g^1 \wedge g^2+g^3
\wedge g^4)\,,
\end{equation}
the $(2,1)$ form reads
\begin{equation} \label{eq:singular21}
\chi_S=\frac{1}{8 \pi^2}\big(\omega_3-3i \frac{dr}{r} \wedge \omega_2\big)\,.
\end{equation}
Notice that this form is `localized' at small $r$. The normalization chosen is related
to the fact that $\int_{S^3}\omega_3=8\pi^2$, although the final results are
independent of this.

The actual solution of interest is the deformed conifold.
In the basis $(\tau,g^i)$ of~\cite{ks} the metric is diagonal
\begin{equation} \label{eq:defg}
ds_6^2=\frac{1}{2}|S|^{2/3}K(\tau)
\big[\frac{d\tau^2+(g^5)^2}{3K^3(\tau)}+{\rm
cosh}^2\frac{\tau}{2}\, \big((g^3)^2+(g^4)^2 \big)+{\rm
sinh}^2\frac{\tau}{2}\, \big((g^1)^2+(g^2)^2\big) \big]
\end{equation}
where
$$
K(\tau):=\frac{\big({\rm sinh}(2\tau)-2\tau
\big)^{1/3}}{2^{1/3}{\rm sinh}\,\tau}\,.
$$
Note that all the moduli dependence is contained in the single
prefactor $|S|^{2/3}$. For large $\tau$, the relation with the
conical radius is
\begin{equation} \label{eq:rtau}
r^2=\frac{3}{2^{5/3}}|S|^{2/3}\,e^{2 \tau/3}\,.
\end{equation}

The $(2, 1)$ form is now more complicated:
\begin{equation} \label{eq:deformed21}
\chi_S=g^5 \wedge g^3 \wedge g^4+d\big[F(\tau)(g^1 \wedge g^3+g^2
\wedge g^4) \big]-i \,d \big[f(\tau) g^1 \wedge g^2+k(\tau) g^3
\wedge g^4 \big]\,,
\end{equation}
where the functions $F$, $f$ and $k$ were computed in~\cite{ks}:
$$
F(\tau)=\frac{{\rm sinh}\,\tau-\tau}{2{\rm
sinh}\,\tau}\;,\;f(\tau)=\frac{\tau\, {\rm coth}\,\tau-1}{2{\rm
sinh}\,\tau}({\rm cosh}\,\tau-1)\;,
$$
\begin{equation} \label{eq:Ffk}
k(\tau)=\frac{\tau\, {\rm coth}\,\tau-1}{2{\rm sinh}\,\tau}({\rm
cosh}\,\tau+1)\,.
\end{equation}

\subsection{Computation of the moduli space metric}\label{subsec:Gconif}

Next we turn on $N$ units of $F_3$ flux through the deformed
3-cycle $A$ of the conifold; as explained in~\cite{ks} this
generates an $H_3$ flux through the dual $B$ cycle. In the
noncompact model it doesn't matter whether $N$ is positive or
negative, and, indeed we will see that the formulas depend only on
$N^2$. The difference will appear when we embed the configuration
in a compact CY; bulk interactions between the `misaligned' flux
$N<0$ and the orientifold will break supersymmetry. This will be
addressed in the next section.

The configuration might be seen either as the gravity side of the
Dijkgraaf-Vafa duality, or as the end of the duality cascade of
Klebanov-Strassler. In any case, the Calabi-Yau is warped as in
(\ref{eq:metric1}); $e^{-4A}$ may be written as~\cite{ks}
\begin{equation} \label{eq:warpKS}
e^{-4A(\tau)}=2^{2/3}\,\frac{(g_s N
\alpha')^2}{|S|^{4/3}}\,I(\tau)
\end{equation}
where $I(\tau)$ is an integral expression defined in~\cite{ks}. It
is not known how to evaluate it in terms of elementary functions,
but fortunately we will only need its derivative:
\begin{equation} \label{eq:derivA}
\frac{d}{d\tau}\,e^{-4A(\tau)}=-4\times 2^{2/3}\, \frac{(g_s N
\alpha')^2}{|S|^{4/3}}\frac{f+F(k-f)}{({\rm
sinh}\,2\tau-2\tau)^{2/3}}\,.
\end{equation}

Although the form of $\chi_S$ is complicated, surprisingly $\chi_S
\wedge \chi_{\bar S}$ is a total $\tau$-derivative: from
(\ref{eq:deformed21}),
\begin{equation} \label{eq:totalderivchi}
\chi_S \wedge \chi_{\bar S}=-\frac{2i}{64\pi^4} d\tau \wedge (\prod_i g^i)\,
\frac{d}{d\tau}\big[f+F(k-f) \big]\,.
\end{equation}
Integrating by parts, the warped metric reads
$$
G_{S \bar S}= -\frac{2i}{\|\Omega\|^2\,64\pi^4 V_W}\,\big(\int \prod_i g^i
\big)\,\Big [ \int_0 ^{\tau_\Lambda}\,d\tau
\frac{d}{d\tau}\big\{e^{-4A(\tau)}(f+F(k-f)) \big\}+
$$
$$
-\frac{de^{-4A}}{d\tau}\,(f+F(k-f)) \Big]\,.
$$
As explained in section \ref{subsec:geometry}, the noncompact
model is regularized at $\Lambda_0$, which determines
$\tau_\Lambda$ through (\ref{eq:rtau}),
\begin{equation}\label{eq:tauL}
\tau_\Lambda=\frac{3}{2}{\rm log}\frac{2^{5/3}}{3}+{\rm
log}\frac{\Lambda_0^3}{|S|}\,.
\end{equation}
We only have to integrate the last term in $G_{S \bar S}$.
Fortunately the integrand decays rapidly; its maximum is $0.03$ at
$\tau \approx 2.5$ and already at $\tau \approx 10$, its value is
$5 \times 10^{-5}$. Since $\tau_\Lambda \gg 1$, with negligible
error we may extend the integral to infinity, giving
$$
\int_0 ^{\infty} d\tau\,\frac{de^{-4A}}{d\tau}\,(f+F(k-f)) \approx
0.093\,\times (-4\times 2^{2/3})\, \frac{(g_s N
\alpha')^2}{|S|^{4/3}}\,.
$$

Putting together these results, we finally obtain the explicit
expression for the metric:
\begin{equation} \label{eq:Gexact}
G_{S \bar S}=-\frac{i}{\|\Omega\|^2 \pi V_W}\Big[c\, {\rm
log}\,\frac{\Lambda_0^3}{|S|}+(8 \times 2^{2/3} \times
0.093)\,\frac{(g_s N \alpha')^2}{|S|^{4/3}} \Big]\,.
\end{equation}
The volume of the base $T^{1,1}$ contributes a factor $64 \pi^3$
and, as before, $c$ is the universal Kahler modulus.

The first term in (\ref{eq:Gexact}) is the usual one, determined
by special geometry and interpreted as integrating out BPS $D3$
branes wrapping the $A$ cycle. The second term is the new
contribution; such a term could not appear in $\mathcal N=2$
compactification, both on mathematical grounds
\cite{Douglas:2005hq} and because loop effects of massless
particles cannot lead to this type of power-like divergence.
However it is a natural consequence of warping in $\mathcal N=1$,
and also has a suggestive interpretation in the dual gauge theory,
as we discuss later.

Notice that at small enough $|S|$, the second term will dominate.
Since it is singular at $S=0$, one should ask whether it is valid in
this regime.  We will examine the consistency condition in
supergravity in subsection \ref{subsec:closer}, concluding that for
$g_s N >> 1$ (the standard supergravity regime) this is valid all the
way down to $S=0$.

\subsection{Gluing the conifold to a compact CY}\label{subsec:gluing}

The advantage of using formula (\ref{eq:Gwarp1}) to compute $G_{\alpha
\bar \beta}$ close to the conifold point ($S \to 0$) is that it is
insensitive to how we patch the noncompact conifold into the
Calabi-Yau. Indeed, in the previous two subsections we saw that the
form $\chi_S$ is localized on the vanishing cycle, so the dominant
contribution to the metric will come from this region. However, since
we are trying to compute a small effective potential and supersymmetry
breaking scale, we might worry that small corrections coming from the
bulk or the patching prescription might qualitatively change the results.

The basic argument that these do not matter, is that the new term
$|S|^{-4/3}$ in \eq{Gexact} grows more quickly as $S\rightarrow 0$ than
any possible bulk term.  At this point we cannot show this for all possible
bulk solutions.  But we do know it for the original unwarped bulk solutions
described by special geometry -- as argued in \cite{Douglas:2005hq,Eguchi:2005eh},
these moduli space metrics can have at most logarithmic divergences, as in
the $c$ term in \eq{Gexact}.  This comes from the integral down to small $r$ and
will not get contributions from elsewhere in the bulk.
Then, if we can argue that the enhancement due to warping is always of the
general form described by \eq{Gwarp0}, we will know that possible bulk
contributions will be subleading to the conifold contribution we computed.

Let us see how \eq{Gwarp0} works for our explicit example.
Volume integrals of the warp factor in the deformed conifold are
hard to make so, for simplicity, we discuss the case of a singular
conifold, which gives the right intuitive picture. We should point
out that even without warping, introducing a cutoff $\Lambda_0$
will make the conifold volume and the bulk volume depend on
$\Lambda_0$ and $S$:
\begin{eqnarray}
V_{CY}&=&V_{bulk}(\Lambda_0, |S|)+{\rm
vol}(T^{1,1})\,\int_{|S|^{1/3}}^{\Lambda_0}\,dr\,r^5\\&=&V_{bulk}(\Lambda_0,
|S|)+\frac{16\pi^3}{27}\,\frac{1}{6}(\Lambda_0^6-|S|^2)\,.
\end{eqnarray}
As we discussed earlier, it is natural to define the total
unwarped volume to be independent of the complex structure, in
which case the bulk volume should have a dependence $\sim |S|^2$
to cancel the conifold contribution.

In the singular conifold, the warp factor generated by $N$
fractional $D3$ branes is~\cite{ks}
\begin{equation} \label{eq:singularA}
e^{-4A(r)}=c+ \frac{81}{8}(g_s N \alpha')^2\,\frac{{\rm
log}\,(r/|S|^{1/3})}{r^4}\,.
\end{equation}
From the gauge theory point of view,
the constant zero mode $c$ corresponds to a dimension 8
operator, which arises naturally as a
correction in the Born-Infeld action. It depends on the amount of
flux turned on, but we can neglect it for large cutoff. On the
other hand, $c_2$ is just the `bare' contribution to the warp
factor, in agreement with the general discussion of subsection
\ref{subsec:zeromodes}.

Now, replacing the value of $c_2$ and (\ref{eq:singular21}) in
(\ref{eq:Gwarp0}),
\begin{equation} \label{eq:Gsemiexact}
G_{S \bar S}=-\frac{i 64 \pi^3}{\|\Omega\|^2 V_W}\Big[c\, {\rm
log}\,\frac{\Lambda_0^3}{|S|}+\frac{81}{32}\,\frac{(g_s N
\alpha')^2}{|S|^{4/3}} \Big]\,.
\end{equation}
Comparing to (\ref{eq:Gexact}), we see that both results have
exactly the same dependence, so our gluing prescription reproduces
the right metric. The ${\rm log} |S|$ coefficients match exactly
because they come from the asymptotic behavior of the conifold,
which is correctly described by the singular conifold. The ones of
$|S|^{4/3}$ have the same order of magnitude, but we don't expect
them to agree. Indeed, this term comes from the contribution of
the deformed 3-cycle, which is described only qualitatively by the
simplified approach of using the singular conifold and introducing
a cutoff at $r=|S|^{1/3}$.

The expression for the vacuum energy from dimensional reduction
is~\cite{gid1}
\begin{equation}
V=\frac{1}{2\kappa_4^2}\,\frac{1}{V_W {\rm Im}\,\tau ({\rm
Im}\,\rho)^3}\frac{1}{\parallel \Omega \parallel^2 V_W}\,G^{S \bar
S} |\partial_S W|^2\,.
\end{equation}
Using the known expression for $G_{S \bar S}$, we obtain
\begin{equation} \label{eq:V3}
V=\frac{1}{2\kappa_{10}^2}\,\frac{g_s}{({\rm
Im}\,\rho)^3}\,\Big[c\, {\rm
log}\,\frac{\Lambda_0^3}{|S|}+c'\,\frac{(\alpha'g_sN)^2}
{|S|^{4/3}} \Big]^{-1} \big|\frac{N}{2\pi i}\,{\rm
log}\,\frac{\Lambda_0^3}{S}+i \frac{\beta^{NS}}{g_S}\big|^2\,.
\end{equation}
To avoid cluttering, we have absorbed order one numerical factors into the constants $c$ and $c'$, although $c$ still denotes
the universal Kahler modulus $c \sim V_{W}^{2/3}$.

\subsection{A closer look into warping effects}\label{subsec:closer}

Near the mouth of the throat, where warping is small, the usual
intuition from special geometry and the deformed conifold is
valid. In particular, in the limit $S \to 0$, the $S^3$ collapses,
its radius being controlled by the factor $|S|^{2/3}$ in the
metric (\ref{eq:defg}).

We may, however, tune the fluxes to get \emph{strong warping}
$e^{-4A} \gg c$. As we now discuss, this changes radically the
picture, even before considering particular models for
supersymmetry breaking. In this subsection we analyze the
supersymmetric $N>0$ case and, in the next one we break
supersymmetry by setting $N<0$.

\begin{figure}[tbf]
\begin{picture}(0,200)%
\includegraphics[scale=1.2]{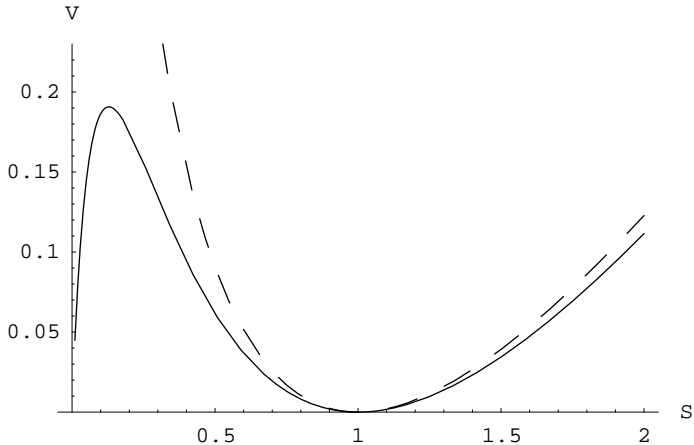}%
\end{picture}%
\caption{\small Behavior of the potential (\ref{eq:V3}) for the
supersymmetric $N>0$ case, with (full line) and without (dashed
line) warping effects. The point $S=1$ is the supersymmetric
vacuum.} \label{fig:conif}
\end{figure}

To begin with, consider a sample potential with and without warping,
plotted in Figure
\ref{fig:conif}.  We have taken order one parameters so that the various
regimes are easily visible on the same plot.

The dashed curve is the potential without the $S^{-4/3}$ correction to
the metric.  It has a minimum given by (\ref{eq:Ssusy}),
where it vanishes, while it goes to infinity at $|S| \rightarrow 0$ and at
$|S| \rightarrow \Lambda_0^3$.  One might have expected that for
$S$ small, the system should become unstable and undergo a
geometric transition.

On the other hand, the behavior of the potential (\eq{V3})
including the warped metric is quite different. At
$$
c\, {\rm
log}\,\frac{\Lambda_0^3}{|S|}\,\sim\,\,\frac{(\alpha'g_sN)^2}{|S|^{4/3}}
$$
the $S^{-4/3}$ starts to dominate; a maximum value is attained and
after that the system starts to roll down to $S=0$!

The reason why this effect was not detected before is that fluxes
break $\mathcal N=2$ softly, so at string tree level the Kahler
metric is still given by special geometry. However, if we want to
analyze the geometric transition in more detail, we have to
consider what happens for $S \to 0$. In this case, the $g_s$
correction of (\ref{eq:V3}) is important, showing that the system
becomes unstable.

Clearly, the supergravity solution is singular at $S=0$. For which
range of small (but finite) $S$ can we trust the supergravity
analysis? To answer this we need to study the curvature of the
background. We consider the `near horizon' limit $\tau \to 0$,
where the largest curvatures may be generated; strong warping
implies the boundary condition $e^{-4A(\tau)} \to 0$ as $\tau \to
\infty$, which is exactly the KS end of the cascade. In this case,
the metric for the warped-deformed conifold
$$
ds_{10}^2= e^{2A(\tau)} \eta_{\mu \nu} dx^\mu
dx^\nu+e^{-2A(\tau)}\,ds_6^2
$$
with $ds_6^2$ given in (\ref{eq:defg}), becomes
\begin{equation}\label{eq:defhorizon}
ds_{10}^2\approx
\frac{1}{2^{1/3}a_0^{1/2}}\,\frac{|S|^{2/3}}{\alpha' g_s N}
\,\eta_{\mu \nu} dx^\mu
dx^\nu+\frac{a_0^{1/2}}{6^{1/3}}\,\alpha'g_s N
\,\big[\frac{d\tau^2}{2}+d\Omega_2^2+d\Omega_3^2\big]\,.
\end{equation}

Here we used the fact that for $\tau \to 0$, the function
$I(\tau)$ introduced in (\ref{eq:warpKS}) behaves as $I(\tau \to
0) \to a_0 \sim 0.7180$~\cite{ks}. Furthermore, we included
explicitly the $S^2$ and $S^3$ at the base of the cone:
\begin{equation}\label{eq:spheres}
d\Omega_2^2=\frac{\tau^2}{2}\big((g^1)^2+(g^2)^2
\big)\,,\,d\Omega_3^2=\frac{1}{2}(g^5)^2+(g^3)^2+(g^4)^2\,.
\end{equation}
The $S^3$ has finite radius, while the $S^2$ collapses, as
expected.

The fact that the $S$ dependence cancels out in
$e^{-2A(\tau)}\,ds_6^2$ is quite striking; this was already
derived in~\cite{ks}, but we would like to point out some of its
consequences. In the strong warping limit, we see that the volume
of the $S^3$ is not proportional to $S$; in particular this
3-cycle \emph{does not} vanish when $S \to 0$! The order of limits
matters and we cannot recover the (strongly) warped deformed
conifold by taking $S\to 0$ in the deformed conifold and then
introducing the warp factor for the singular conifold. The modulus
$S$ no longer parametrizes the size of a cycle in the warped
deformed geometry. Note, however, that not all the dependence on
$S$ of the six-dimensional geometry has disappeared. Indeed,
unlike $r$, $\tau$ is a dimensionless coordinate; cutting off the
conifold at some finite $\tau_\Lambda$ requires both scales
$\Lambda_0$ and $S$, as showed in (\ref{eq:tauL}). Hence, as $S
\to 0$, the throat becomes infinite (even at fixed $\Lambda_0$).
Of course, once $e^{-4A}$ is small enough, the bulk effects become
relevant, cutting off the geometry; but still, this behavior is
very different to the deformed case without warping.

The analysis of the curvature tensor of (\ref{eq:defhorizon}) is
straightforward. A crucial point is that the only dependence on
$S$ is through $\eta_{\mu \nu} dx^\mu dx^\nu$; since the curvature
does not depend on $x^\mu$, defining orthonormal Minkowski
coordinates
\begin{equation}\label{eq:orthonormal}
\tilde x^\mu:=\big(\frac{1}{2^{1/3}a_0^{1/2}}\,\frac{|S|^{2/3}}{
\alpha' g_s N} \big)^{1/2}\,x^\mu\,,
\end{equation}
none of the components of $R^M_{\phantom{1}NRS}$ will depend on
$S$. An explicit computation to order $\tau^2$ gives the scalar
curvature
\begin{equation}\label{eq:R}
R=-\frac{6^{1/3}}{5 \sqrt a_0}\,\frac{1}{\alpha' g_s N
}\big[3(1+20k)-(6+9 k+880k^2)\tau^2 \big]+\mathcal O (\tau^3)
\end{equation}
where $I(\tau)\sim a_0 (1+k\tau^2)$, $k$ being an order one
constant. Therefore, unlike the unwarped case, we can trust the
supergravity approach as long as $g_s N \beta^{NS} \gg 1$, even if
$S \to 0$ (but finite). Incidentally, (\ref{eq:orthonormal})
implies that the time $x^0$ necessary to roll down to $S=0$ tends
to infinity, at fixed orthonormal time $\tilde x^0$.

If the modulus $S$ doesn't have now a geometric interpretation,
what is its meaning? As explained by KS, (\ref{eq:defhorizon}) is
the `supergravity version' of confinement. Since the prefactor
multiplying $\eta_{\mu \nu} dx^\mu dx^\nu$ is finite for $\tau=0$,
Wilson loops will have an area law. Furthermore, we see that the
theory generates dynamically a confinement scale
\begin{equation}\label{eq:Mconf}
M_{conf}^2 \sim \frac{|S|^{2/3}}{\alpha' g_s N}
\end{equation}
controlled by $S$. From this point of view, the previous
noncommutativity of limits is expected: the warped singular
conifold cannot reproduce these nonperturbative effects.

\subsection{Breaking supersymmetry at strong warping}\label{subsec:bulkeffects}

We have finished assembling the necessary tools to understand how
warp effects influence the nonsupersymmetric $N<0$ case. The
unwarped case was considered in~\cite{ag} and summarized in
subsection \ref{subsec:susydef}.

To evaluate the scale of this breaking we need $G_{S \bar S}$,
which was computed explicitly in subsection \ref{subsec:Gconif}.
Integrating (\ref{eq:Gexact}), the Kahler potential for the
conifold, in rigid special geometry, reads
\begin{equation} \label{eq:Kconifw}
K(S, \bar S)=-\frac{i 64
\pi^3}{\|\Omega\|^2V_W}\big[c\,|S|^2\,{\rm
log}\,\frac{\Lambda_0^3}{|S|}+ (72\times 2^{2/3}\times
0.093)\alpha'g_s N  |S|^{2/3} \big]\,.
\end{equation}
We are interested in the regime where the new warp correction dominates, namely
\begin{equation} \label{eq:susycond}
c\,{\rm log}\,\frac{\Lambda_0^3}{|S|} \ll \frac{(\alpha' g_s
N)^2}{|S|^{4/3}}\,.
\end{equation}
From (\ref{eq:Snsusy}) this may be attained by an adequate choice of fluxes $\beta^{NS} \gg g_s N$.

Replacing in (\ref{eq:V2}) the values of $S_{N<\,0}$,
$\partial_S W_{N<\,0}$ and $G_{S \bar S}$ that we found,
\begin{equation} \label{eq:Msusy}
M^4_{susy}:=V_{N<\,0}=\frac{k}{\kappa_4^2}\,\frac{1}{V_W ({\rm
Im}\,\rho)^3g_s}\,\big|\frac{\beta^{NS}}{g_sN}\big|^2 {\rm
exp}\big(-\frac{8\pi}{3}\frac{\beta^{NS}}{g_s|N|}\big)\,\Lambda_0^4\,.
\end{equation}
This has the desired exponential suppression in the semiclassical
limit \\ $\beta^{NS}/g_sN\ \gg 1$. Note that in the limit $V_W \to
\infty$, the orientifold will be far away from the throat and
$V_{N<0} \to 0$, which agrees with the idea that the system is
locally supersymmetric.

From the point of view of the potential (\ref{eq:V3}), the
prescription of~\cite{ag} for the physical vacuum
(\ref{eq:Snsusy}) puts us in an unstable point, rolling directly
to $S=0$! One option would be that the present description, in
terms of a single field $S$ is not valid in strongly warped
regimes. Indeed, in the known holographic descriptions of confined
pure SYM~\cite{ks, malda, ps}, the masses of KK modes (from
dimensional reduction on the conifold) are comparable to the
glueball mass. Including these fields is not a simple task,
requiring, in particular, a better understanding of the Green's
functions on the deformed conifold.

Here we briefly discuss another option, namely that the breaking
of the no-scale structure (due to the absence of supersymmetry)
may stabilize the vacuum. Indeed, as the analysis of~\cite{iss}
suggests, metastable vacua in general require two scales, one
generated by the gauge theory, and another coming from UV effects
(in their case, the small mass $m$ for quarks).  Our discussion so
far has no analog of this second scale.

To begin a full discussion, one would have to incorporate the various
ingredients of moduli stabilization discussed in \cite{Kachru:2003aw},
including stabilization of the dilaton and the complex structure
moduli other than $S$, and breaking of no-scale structure and
stabilization of K\"ahler moduli due to stringy and quantum
corrections which depend on the these moduli.  We now assume that this
has been done in some way which does not affect the physics in the
throat, and discuss the remaining physics in the throat after integrating
these modes out, using the supergravity potential
\begin{equation}
V=\kappa_4^2 e^K\Big[G^{S \bar S}|D_S W|^2-3|W|^2 \Big]
\end{equation}
Actually this expression would only be exact in a limit in which
the other moduli were infinitely massive; otherwise it will receive corrections
from cross-coupling between the other moduli and $S$.  However, one can
easily state conditions under which these effects will not
qualitatively affect the results, so we neglect this.

Now, taking the anti-self-dual flux configuration, an important point is that
the dual period $\partial \mathcal F / \partial S$ does not vanish
in the limit $S \to 0$. Writing
$$
\int_B\Omega ={S\over 2\pi i} {\rm log} {\Lambda_0 ^ 3\over S} +\Pi_0\,,
$$
then
\begin{equation}
W = {N\over 2\pi i} S
 \big({\rm log}\frac{\Lambda_0^3}{S}+1 \big)+N\Pi_0+\frac{i}{g_s} \beta^{NS}S
\end{equation}
and the condition for having a minimum at small $S$ is
\begin{equation}
S^{1/3}\,{\rm log}\,\frac{\Lambda_0}{S^{1/3}} \approx \frac{2\pi
c' \,\Pi_0}{\int e^{-4A}\Omega \wedge \overline \Omega}\,(\alpha'
g_s N)^2 \,.
\end{equation}
Here we are using the notation of Eq. (\ref{eq:V3}).

Typically,
$$
\frac{2\pi \Pi_0}{\int e^{-4A}\Omega \wedge \overline \Omega} \sim V_W^{-1/2}
$$
so by choosing a bulk volume $V_W^{1/2} \gg \alpha'^2 g_s N
\beta^{NS}$, the modulus is stabilized at a parametrically small
(though no longer exponentially small) scale.  The vacuum energy
here is of the order
$$
V_{min} \approx \frac{1}{2\kappa_{10}^2}\frac{g_s}{{\rm Im
\rho}^3}\,\frac{|\Pi_0|^2}{\int e^{-4A}\Omega \wedge \overline
\Omega}\,.
$$

Since the natural scale of the potential away from the $S\to 0$ limit is set
by $\beta^{NS}/g_sN$, the height and breath of the barrier separating this
minimum from the true vacuum scale in the same way.  It's worth
mentioning at this point that in the GKP conifold setup the parameter
choices leading to a controllable hierarchy are $1 << g_s N,$ so that
the supergravity approximation is reliable at the tip, and $1 << \beta^{NS}/g_sN$,
which in the supersymmetric GKP setup sets the scale of the hierarchy.
These are precisely the same relations which yield a reliable
metastable vacuum here.

Finally, once one has found a stable vacuum from the point of view
of the $\mathcal N=1$ effective Lagrangian, one needs to ask
whether other effects could destabilize it, in particular whether
a KK mode which was dropped in deriving the Lagrangian could go
tachyonic. The basic answer to this question is that, since there
is a limit in which the throat solution would have been $\mathcal
N=1$ supersymmetric had not that supersymmetry been projected out
by the orientifolding, it will satisfy the constraints of this
$\mathcal N=1$ supersymmetry, up to small corrections. Thus, one
can restrict attention to the light modes in the $\mathcal N=1$
effective Lagrangian, and see whether the new couplings introduced
at this point destabilize any of them; massive KK modes will be
stable since the original KS solution was stable.

\section{The dual gauge theory} \label{sec:gauge}

The discussion we just gave should be valid for $g_s N >> 1$.  For
small $g_s N$, we would expect a description in terms of the gauge
theory on the wrapped anti-D5 branes to be more appropriate.
We do not know how such a description would work in detail, but
we can make the following comments on the problem.

Let us start by considering the embedding of the conifold with
anti-self-dual flux into an $\mathcal N=2$ compactification.
There, the gauge theory under discussion is {\em the same} as the
gauge theory usually invoked in this duality, namely the
$U(N_1)\times U(N_2)$ supersymmetric gauge theory of \cite{ks} in
the UV, undergoing a ``cascade'' down to pure $U(N)$ super
Yang-Mills theory. This theory has $N$ supersymmetric vacua, and
we recover the standard discussion, with the sole change being the
sign of the RR fluxes and the identification of the unbroken
$\mathcal N=1$ subalgebra in $\mathcal N=2$.

As we saw in the supergravity analysis, it seems very plausible
that the essential phenomenon is a misalignment of the $\mathcal
N=1$ supersymmetries preserved by the bulk and by the antibrane.
To describe this in gauge theory terms, we might try to identify
the action of bulk $\mathcal N=2$ supersymmetry on the gauge
theory, and the $\mathcal N=2$ stress tensor multiplet, which
would couple to the $d=4$, $\mathcal N=2$ supergravity obtained by
KK reduction.  The difference between the D5 and anti-D5 theories
then arises when we do the orientifold projection, obtaining a
$d=4$, $\mathcal N=1$ supergravity.  Whereas for the D5 theory, we
couple the $\mathcal N=1$ stress tensor multiplet to $\mathcal
N=1$ supergravity, for the anti D5-brane we would instead couple
to the {\em broken} $\mathcal N=1$ subalgebra of $\mathcal N=2$.

This idea is simple to realize in the case of branes embedded in
flat space.  Consider for example the world-volume theory of $N$
D3-branes; it is ${\mathcal N}=4$ super Yang-Mills with $16$
linearly realized supersymmetries.  It also has $16$ nonlinearly
realized supersymmetries, the constant shifts of the diagonal
components of the gauginos.  An analog of the theory under
discussion is obtained by truncating this to a linearly realized
$\mathcal N=1$ and a nonlinearly realized $\mathcal N=1$.  Thus,
the antibrane couples to the $\mathcal N=1$ gravitino, not through
the standard supercurrent, but through the gaugino.

This leads to spontaneous supersymmetry breaking, at a scale controlled
by the antibrane tension.  However, it is not obvious how strong coupling
effects could lower this scale.  Naively, since the supersymmetry breaking
is all in the coupling to the $U(1)$ sector, the nonabelian Yang-Mills
sector does not seem to play any role.  However, the sectors could
be coupled by higher dimension operators, so this conclusion is probably
too quick.

According to the usual discussions of the AdS/CFT correspondence,
the $\mathcal N=2$ supersymmetry of the underlying string
background, is reflected in the $\mathcal N=1$ superconformal
symmetry of the gauge theory. Thus, the idea would be to couple
the gravitino of $\mathcal N=1$ supergravity, not to the standard
supercurrent, but to the superconformal current of the gauge
theory.

Unfortunately, this idea is not consistent as it stands, as the
superconformal symmetry in these gauge theories is explicitly broken
by quantum effects (the beta function is non-zero) and we cannot gauge
an explicitly broken symmetry.  Nevertheless it might be correct if
a suitable compensator field is present in the bulk theory.

\subsection{Effective potential}

Granting that there is a microscopic definition of the theory as a
gauge theory coupled to $\mathcal N=1$ supergravity, we next ask
whether the effective potential we have derived and justified at $g_s
N >>1$, should be expected to give a good qualitative description
for small $g_s N$.  As we commented earlier, even in the supersymmetric
vacua the precise interpretation of this type of effective action is not
entirely clear, so we limit ourselves to questions about vacuum energy,
supersymmetry breaking and stability.

We begin with the $|S|^{2/3}$ term in the K\"ahler
potential.  It is amusing and perhaps significant that
such a term was already suggested in the pioneering work of
Veneziano and Yankielowicz on  pure $SU(N)$ SYM \cite{vy}.
The argument there was that the gaugino bilinear $S$,
being a composite field, does not have the canonical dimension of
a scalar field.  At weak coupling, its dimension should be close
to that of a fermion bilinear in free field theory, namely
$[S]=2[\lambda]=3$. On the other hand,
a $d^4\theta$ kinetic term should have dimension $2$.
If we are not allowed any dimensionful constants, this forces
\begin{equation} \label{eq:Kgauge}
K(\mathcal S \bar{\mathcal S})=\alpha(\bar{\mathcal S} \mathcal
S)^{1/3}\,
\end{equation}
for some numerical constant $\alpha$.  This
precisely matches the new term coming from warping in (\ref{eq:Kconifw})!

Unfortunately, for $S\rightarrow 0$ the gauge theory is strongly
coupled, and it is not known how to compute the K\"ahler potential in
this regime.  On general grounds one would expect corrections controlled by
the dynamical scale $\Lambda$.  While it is true that $\Lambda$ does
not appear explicitly in the superpotential, only emerging upon solving
for the vacuum, there is no obvious reason that the K\"ahler potential
should work the same way.  Thus at this point we can not say we have
strong evidence for such a term at weak coupling, although it is
certainly a very suggestive coincidence.

In any case, if we accept that the theory breaks supersymmetry at
the dynamical scale, the claim that the metric $G_{S\bar S} \sim
|S|^{-\alpha}$ for some $\alpha>0$ would seem to be a very natural
way to describe this in an $\mathcal N=1$ effective Lagrangian. It
might not be inevitable, as one can also imagine inverse powers of
$\Lambda$ playing this role.  However, this would violate the
general principle that nonperturbative effects should vanish in
the weak coupling limit $\Lambda\rightarrow 0$, so it seems a
reasonable hypothesis that such effects are not present.

But, as we saw in our explicit example, any structure in which the
vacuum energy is warped down by a power of $S$, leads directly to a
potential with a zero energy minimum at $S=0$.
We discussed how in a string theory compactification this might be
prevented by bulk effects.  But in the gauge theory limit, such
effects would presumably be absent, so the result would be a theory
which rolls down to $S=0$.

Could there be another supersymmetric vacuum at $S=0$?
A suggestion that super Yang-Mills theory has additional vacua
at $S=0$ was made in \cite{Kovner:1997im}, however at present this
is not believed to be the case.

One straightforward way to reconcile these claims is if the
effects we are discussing, in particular the correction to the
K\"ahler potential and the corresponding lowering of the
supersymmetry breaking scale, are not present at small $g_s N$.
Now some brane-antibrane
realizations of supersymmetry breaking, for example
\cite{Bena:2006rg,Giveon:2007fk}, lead to a non-trivial
phase structure, and it might be the case here.  However,
in these realizations, the supersymmetry breaking vacuum exists
at weak coupling, and disappears at strong coupling, so the opposite
claim might be surprising.

It also seems possible to us that while this effective field
theory is qualitatively valid, the configuration rolling down to
$S=0$ is not a conventional vacuum.  This is true in the
supergravity limit, as the value of $S$ controls the warp factor
in $d=4$, so that $S=0$ cannot be realized.  One can still imagine
solutions in which $S$ rolls to zero, but these are essentially
dynamical.  In particular, since the warp factor multiplies the
$g_{00}$ component of the metric, the time evolution is very
different than the flat space evolution in such a potential.
As we explained in the discussion below \eq{R}, this suggests that
the minimum $S=0$ is not reached in finite physical time.
We intend to study the physics of this more carefully
in a future work.

\vskip 4mm

\subsection*{Acknowledgments} We would
like to thank F. Denef, D.-E. Diaconescu, R. Essig, J. Juknevich,
I. Klebanov, S. Klevtsov, D. Kutasov, S. Lukic, D. Melnikov, A.
Nacif, S. Ramanujam, N. Seiberg, K. Sinha, M. Strassler, S.
Thomas, K. van den Broek, G. Veneziano and S. Yankielowicz for
useful comments and discussions.  We also thank S. Kachru and S.
Giddings for valuable comments on the manuscript.

This research was supported by DOE grant DE-FG02-96ER40959.

\end{document}